Perspective

# Manufacturing carbon nanotube transistors using lift-off process: limitations and prospects

Xilong Gao[1,2], Jia Si[2]*, and Zhiyong Zhang[2]

[1]School of integrated Circuits, Beijing University of Posts and Telecommunications, Beijing 100876, China

[2]Key Laboratory for the Physics and Chemistry of Nanodevices and Center for Carbon-based Electronics, School of Electronics, Peking University, Beijing 100871, China

*E-mail: sijia@pku.edu.cn

ORCID: Jia Si 0000-0003-0737-4905; Zhiyong Zhang 0000-0003-1622-3447

**Abstract**

Carbon nanotube field-effect transistors (CNT FETs) are regarded as promising candidates for next-generation energy-efficient computing systems. While research has employed the lift-off process to demonstrate the performance of CNT FETs, this method now poses challenges for enhancing individual FET performance and is not suitable for scalable fabrication. In this paper, we summarize the limitations of the lift-off process and point out that future advancements in manufacturing techniques should prioritize the development of etching processes.

As modern integrated circuits (ICs) advance toward improving performance/integration density, and lowering power consumption, the quest for new semiconductor that can surpass the limitations of traditional silicon-based options has become increasingly crucial [1-2]. One of the most promising alternatives is semiconducting single-walled carbon nanotube, which offers exceptional carrier mobility, high saturation velocity, potential for miniaturization, feasible of complementary metal-oxide-semiconductor (CMOS) logic, and compatible to wafer-scale fabrication [3-5]. CNT CMOS have then been considered as the building blocks for future energy-efficient semiconductor applications, particularly as we approach the 1nm node [6].

Significant progress has been achieved in exploring the intrinsic performance of CNT FETs through the lift-off process, a common laboratory technique, as shown in Fig. 1a. Photoresist (PR) layer is spin-coated and developed, followed by film deposition. After immersion this structure in a solution allows for the removal of the PR, enabling the film above it to be lifted off. This straightforward process enables a gate length of 5 nm with a single CNT [7]; however, it has limitations in optimizing performance for FETs based on aligned CNT (ACNTs) arrays, which are the optimal configuration for generating sufficient driving current [8]. One limitation is that PR typically cannot withstand temperatures exceeding 200 °C, which compromises the quality of the dielectric layer. This layer requires high growth temperature during atomic layer deposition (ALD) and post deposition annealing (PDA) to eliminate defects, fixed charges, and interface states within the gate stack. High temperatures can accelerate reaction rates, enhance the efficiency of precursor decomposition, and promote the formation of stable, dense compounds, which helps to minimize fixed charges. Furthermore, elevated temperatures increase atomic activity, aiding in the removal of irregular lattice structures and alleviating internal stress within the material, thereby further decreasing defect formation. Recently, a gate dielectric on ACNTs was achieved through the ALD of $HfO_2$ at 90°C, yielding a minimum interface state density of $6.1\times10^{11}$ $cm^{-2}$ $eV^{-1}$ [9]. It is anticipated that employing high-temperature ALD processes (exceeding 250°C) combined with PDA could meet the requirements for advanced MOSFETs, targeting an interface state density close to $1\times10^{11}$ $cm^{-2}$ $eV^{-1}$. Additionally, during the final stage of the lift-off process, the detachment of films can result in random deformations, such as warped edges and burrs (Fig. 2a and b), which can lead to severe variations and reliability issues. Furthermore, CNTs are often exposed to PR during the lift-off process, causing chemical contamination and displacement (Fig. 2c and d).

The lift-off process also restricts innovations in device structure. As shown in Fig. 1b, the typical lift-off process begins with defining the source and drain regions on the channel area, followed by dielectric growth on the channel and source-drain electrodes, and finally forming a gate on the dielectric. This sequence results in substantial overlap between the gate and source/drain, which leads to increased parasitic capacitance. For N-type CNT FETs, low work function metals, such as scandium (Sc) are employed; however, during the fabrication process, oxygen or water vapor can easily infiltrate the interface between the Sc contacts and the ACNTs, potentially forming an oxide layer beneath the Sc (see Fig. 2e) and leading to reduced contact length and device failure [10].

Beyond the limitations in performance and structure, the lift-off process poses engineering challenges. As transistor sizes shrink, the demand for precise patterning technologies increases, requiring finely patterned graphics. The lift-off process becomes particularly problematic at scales below 100 nm, where uniformity can be compromised by film residues (see Fig. 2f and g). Additionally, the technique often struggles with wafer-scale integration (see Fig. 2h), resulting in poor yield. Recently, a tensor processor unit utilizing 3000 CNT FETs showcased the best scalability of the lift-off process [11]. Further improvements in scalability require the innovations in new process.

In summary, the lift-off process imposes limitations on performance, device structure, and scalability in the development of CNT technology. Future advancements in manufacturing techniques should prioritize the development of etching processes compatible with very-large-scale integration (VLSI) technologies. However, this presents challenges, as plasma-induced physical bombardment can easily damage CNTs. Therefore, exploring damage-free etching methods, such as designing etching stop layers or implementing mild atomic layer etching (ALE) techniques could be effective strategies. The etching stop layer is designed to exhibit high etch selectivity, effectively protecting CNTs from damage during the etching process. Its thickness must be carefully considered to ensure sufficient protection and its etching byproducts should be easy to remove in subsequent steps [12]. ALE is a precise material removal technique that alternates between chemical reactions and physical removal, eliminating material one atomic layer at a time. By carefully adjusting parameters such as precursor types, plasma power, pressure, gas composition, and temperature, it is possible to achieve highly precise control. This makes ALE one of the most gentle technologies for manufacturing nanostructures [13]. Additionally, new etching recipes tailored specifically for metals used in CNT integrated circuits, such as Pd, Sc, and Y, must be developed [14].

In conclusion, the advancement of CNT integrated circuits must be hindered by the lift-off process, underscoring the urgent need for manufacturing innovations to bridge the gap between laboratory research and industrial application.

**Statements and Declarations**

**Ethics statement for the use of human and animal subjects (may require consent to participate and consent to publish for human subjects).** Not applicable.
**Consent for publication** Not applicable.
**Competing Interest** No.
**Author's Contribution** Jia Si supervised this project, Xilong Gao wrote the manuscript, Jia Si and Zhiyong Zhang revised the manuscript.
**Funding**
This work was supported by the National Natural Science Foundation of China (No.62274006 and 62225101), the National Key Research and Development Program of China (2022YFB4401601), Guangdong Major Project of Basic Research (Grant No. 2021B0301030003), Jihua Laboratory (Project No. X210141TL210), and Research Project: CNT-DSP-2022.
**Data availability** Data sharing not applicable to this article as no datasets were generated or analysed during the current study.

(a)

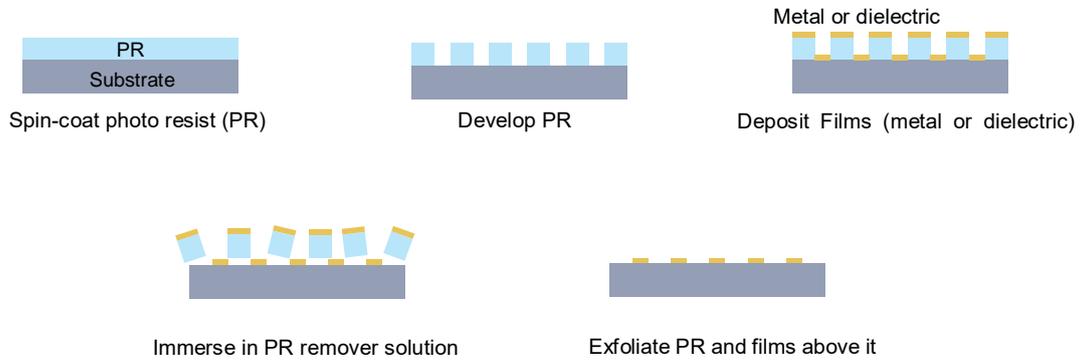

(b)

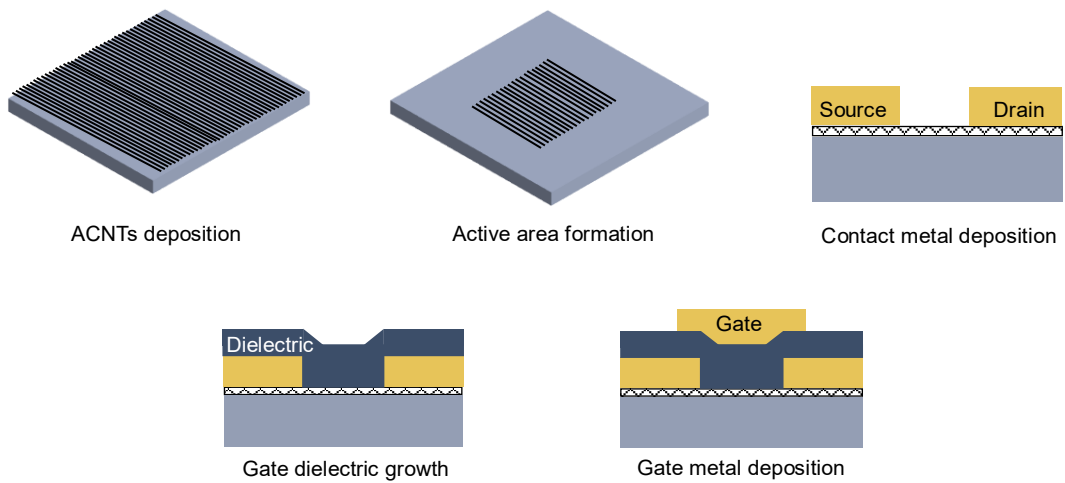

**Fig. 1** Manufacturing process and flow of CNT FETs. **a** Schematic diagram illustrating the typical lift-off process. **b** Schematic diagram of a CNT-based device fabricated using the lift-off process

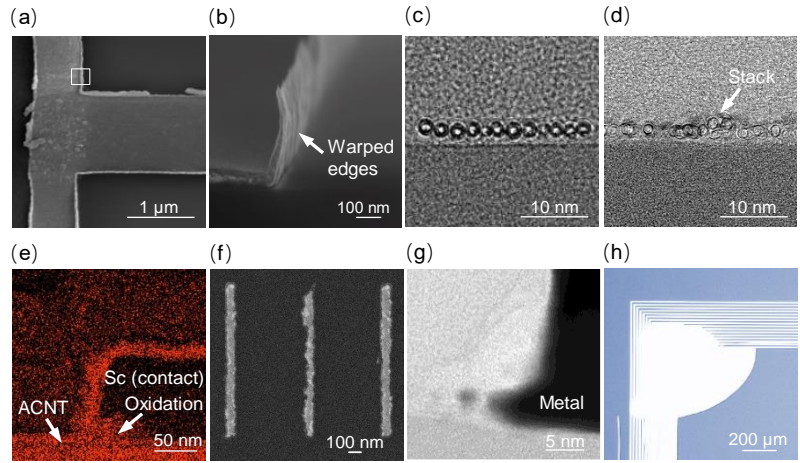

**Fig. 2** Non-idealities during the lift-off fabrication processes of CNT FETs and circuits. **a** SEM image of warped edges. **b** Enlarged sectional view of the edges in panel **a**. **c** TEM image of pristine ACNT arrays, demonstrating a uniform single-layer structure. **d** TEM image of ACNT arrays obtained after the lift-off process, showing distinct bundles and stacks. **e** Oxidation of the n-type contact metal Sc, resulting in a reduced contact length. **f** Non-uniform metal lines produced during the lift-off process. **g** Beak-like metal residuals. **h** Failed to exfoliate metal lines on an 8-inch wafer during the lift-off procedure

**Authors information**

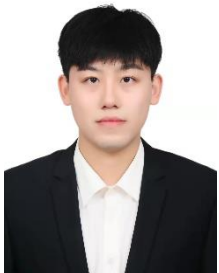

**Xilong Gao** received his Bachelor's degree in Electronic Science and technology from Beijing University of Chemical and Technology in 2022. He is currently pursuing a master's degree in Next-Generation Electronic Information at Beijing University of Posts and Telecommunications. His research interest mainly focuses on the fabrication of high-performance field-effect transistors based on carbon nanotubes.

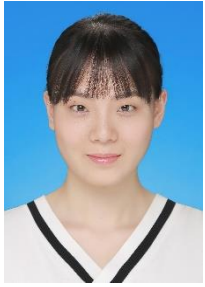

**Jia Si** obtained a doctoral degree from the school of electronics, Peking University in 2019. She is now an assistant research professor at the carbon-based electronics research center in Peking University. Her main research interests include new principle electronic devices, three-dimensional monolithic integration systems, and energy-efficient carbon-based computing circuits.

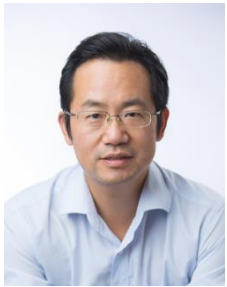

**Zhiyong Zhang** is now a professor at the school of electronics, Peking University, he is the director of Key Laboratory for the Physics and Chemistry of Nanodevices and deputy director of the Center for Carbon-based Electronics at Peking University. He mainly engaged in research on carbon-based nanoelectronics, exploring CMOS integrated circuits, sensors, and other new information device technologies based on carbon nanotubes, and promoting the practical development of carbon-based information device technology.